 \documentclass[preprint,12pt]{elsarticle}

\usepackage{amssymb}
\usepackage{graphicx}
\usepackage{epstopdf, epsfig}
\usepackage{lineno,hyperref}
\usepackage{booktabs} 
\usepackage{array} 
\usepackage{amsmath}
\usepackage{mathtools}
\usepackage{graphicx,caption,subcaption,color}

\journal{Computers \& Fluids}

\begin{document}

\begin{frontmatter}

\title{Isotropy and spurious currents in pseudo-potential multiphase lattice Boltzmann models}

\author[label1]{Cheng Peng\corref{cor1}}
\address[label1]{Department of Energy and Mineral Engineering and EMS Energy Institute,
The Pennsylvania State University,
University Park, PA 16802, USA}

\cortext[cor1]{Corresponding author}

\ead{czp341@psu.edu}

\author[label1]{Luis F. Ayala}

\author[label2]{Orlando M. Ayala}
\address[label2]{111A Kaufman Hall, Department of Engineering Technology, Old Dominion University, Norfolk, VA, 23529, USA}

\author[label3,label4]{Lian-Ping Wang}
\address[label3]{126 Spencer Lab, Department of Mechanical Engineering, University of Delaware, Newark, DE, 19711, USA}
\address[label4]{Department of Mechanics and Aerospace Engineering, Southern University of Science and Technology,
Shenzhen 518055, Guangdong, China}

\begin{abstract}
The spurious currents observed in multiphase flow simulations with pseudo-potential lattice Boltzmann (LB) models are usually understood to be the result of  the lack of isotropy of the model-generated interaction force between phases. Remedies have been proposed to utilize larger stencils to compute the interaction force with higher orders of isotropy. In this short communication, we point out the incompleteness in the current understanding and propose a new consistent implementation to more effectively suppress the spurious currents. We also demonstrate theoretically that certain low-level spurious currents cannot be eliminated by increasing isotropy if the local hydrostatic balance inside the diffuse interface is not established in the LB models.
\end{abstract}


\end{frontmatter}

\section{Introduction}
The pseudo-potential multiphase lattice Boltzmann (LB) models, (also known as the Shan-Chen models~\cite{shan1993lattice,shan1994simulation}, have been widely applied to study a wide range of multiphase flow problems. Despite their successes, the existence of non-physical flux around a steady and static two-phase interface, known as spurious currents, still plagues most multiphase applications and remains an unresolved problem. 

Many efforts were made to explore the origin of these spurious currents and to suppress them. Those efforts are comprehensively reviewed in the literature~\cite{chen2014critical,li2016lattice}. Wagner pointed out that spurious currents were introduced by an incompatible discretization of the interaction force in the two-phase LB model~\cite{wagner2003origin}. Shan~\cite{shan2006analysis} and Li and Fischer~\cite{lee2006eliminating} both realized that when the discretization schemes of the interaction force in the multiphase LB models lack isotropy, spurious currents would emerge. Yuan and Schaefer reported that certain equation of state (EOS) could potentially reduce the level of spurious currents~\cite{yuan2006equations}. Yu and Fan found that, compared to the single-relaxation-time (SRT) LB models, the multiple-relaxation-time (MRT) LB models could be used to suppress the spurious currents by tuning the relaxation parameters irrelevant to the Navier-Stokes equation~\cite{yu2010multirelaxation}. Guo {\it et al.} concluded that spurious currents were inevitable due to intrinsic imbalance of interaction force and the density gradient in the pseudo-potential multiphase LB models~\cite{guo2011force}. Mattila {\it et al.} suspected that the existence of spurious currents could be associated with the second-order accuracy of LB models due to the trapezoidal time-integration scheme, they therefore proposed to use higher-order LB models to suppress the spurious currents.

\section{Shan's improvement and its incompleteness}
Among all these explanations, a rather well-known explanation of the origin of spurious currents in the psuedo-potential LB models was given by Shan, who realized that the high-order terms in the Taylor series of the interaction force, {\it i.e.}, the interaction force exerted on one phase due to the existence of another phase around, lack the required isotropy~\cite{shan2006analysis}. The interaction force ${\bf F}$ in the pseudo-potential multiphase LB models is computed as
\begin{equation}
    {\bf F}\left({\bf x},t\right) = -G\psi\left({\bf x},t\right)\sum_{\alpha}w_{\alpha}\left(|{\bf e}_{\alpha}|\right)\psi\left({\bf x}+{\bf e}_{\alpha}\delta t,t\right){\bf e}_{\alpha},
    \label{eq:Shanchenmodel}
\end{equation}
where $G$ is a parameter measuring the intensity of the interaction, $\psi$ is the field potential that is a function of local fluid density $\rho$, ${\bf e}_{\alpha}$ is the vector stencil employed in the computation of interaction force, which is not necessarily the same as the discrete velocity set in LB models, $w_{\alpha}$ is the corresponding weighting factor and $\delta t$ is the time step size.  Eq.~(\ref{eq:Shanchenmodel}) can be expanded in terms of a Taylor series at ${\bf x}$ and $t$, using tensor notations, as
\begin{equation}
\begin{split}
    &F_{i} = -G\psi\sum_{\alpha}w_{\alpha}e_{\alpha i}\left[\psi + \delta t e_{\alpha j}\partial_{j}\psi \right .\\
    &\left .+ \frac{1}{2}\delta t^2 e_{\alpha j}e_{\alpha k}\partial_{j}\partial_{k}\psi + \frac{1}{6}\delta t^{3}e_{\alpha j}e_{\alpha k}e_{\alpha l}\partial_{j}\partial_{k}\partial_{l}\psi+\cdots\right].
    \end{split}
    \label{eq:Taylorseries}
\end{equation}
Except the first term, each term in the above Taylor series contains a part 
$\sum_{\alpha}w_{\alpha}\underbrace{e_{\alpha i}e_{\alpha j}e_{\alpha k}e_{\alpha l}\cdots}_{n~e_{\alpha}}$, which is a $n$th order tensor. Shan pointed out that spurious currents were originated from the lack of complete isotropy 
of these high-order tensors, when the number of ${\bf e}_{\alpha}$ is finite. As a remedy, Shan employed larger stencils to compute ${F}_{i}$, which allowed additional tensors to be isotropic and increased the order of isotropy in the computed interaction force $F_{i}$. For example, in two space 
dimensions, the highest order of the isotropy realizable can be increased from fourth with the velocity stencil shown in Fig.~\ref{fig:stencils}a to eighth with the stencil in Fig.~\ref{fig:stencils}c. 

In this short communication, we would like to point out the incompleteness in Shan's recommended remedy and its implementation. In fact,  there is a second aspect in terms of isotropy that has usually been ignored but plays an important role in inducing spurious currents. To explain this, let us recall the algorithm of the LB method
\begin{equation}
\begin{split}
   &f_{\beta}\left({\bf x},t+\delta t\right) = f_{\beta}\left({\bf x}-{\bf c}\delta t,t\right)\\
   &-\frac{1}{\tau}\left[f_{\beta}\left({\bf x}-{\bf c}\delta t,t\right)-f_{\beta}^{(eq)}\left({\bf x}-{\bf c}\delta t,t\right)\right]+\phi_{\beta}\left({\bf x}-{\bf c}\delta t,t\right)
   \end{split}
    \label{eq:LBE}
\end{equation}
where ${\bf c}_{\beta}$ is the discrete velocity set in LB model that may be different from ${\bf e}_{\alpha}$, $\tau$ is the relaxation time. 
The equilibrium distribution $f_{\beta}^{(eq)}$ and the forcing function $\phi_{\beta}$ are defined as
\begin{equation}
\begin{split}
    &f_{\beta}^{(eq)} = \rho w_{\beta}\left[1+\frac{{\bf c}\cdot{\bf u}}{c_s^2} + \frac{\left({\bf c}\cdot{\bf u}\right)^2}{2c_s^4}-\frac{{\bf u}\cdot{\bf u}}{2c_s^2}\right],\\
    &\phi_{\beta} = \left(1-\frac{1}{2\tau}\right)w_{\beta}\left[\frac{{\bf c}-{\bf u}}{c_s^2}+\frac{\left({\bf c}\cdot{\bf u}\right){\bf c}}{c_s^4}\right]\cdot{{\bf F}}\delta t,
    \end{split}
    \label{eq:feq}
\end{equation}
where $c_s$ is the speed of sound, which is an input parameter in a specific LB model based on numerical quadrature requirements. The forcing function in Eq.~(\ref{eq:feq}) is the one proposed by Guo~\cite{guo2002discrete}, which ensures a second-order accurate body force term in the reproduced Navier-Stoke equation. We further assume a zero velocity field ${\bf u}(t) = 0$ is reached at the current time $t$, then Eq.~(\ref{eq:feq}) is simplified as
\begin{equation}
    f_{\beta}^{(eq)} = \rho w_{\beta},~~~\phi_{\beta} = \left(1-\frac{1}{2\tau}\right) w_{\beta}\frac{{\bf c}\cdot{\bf F}}{c_s^2}\delta t,
    \label{eq:feqsimpler}
\end{equation}
Finally, for demonstration purposes, a special case $\tau = 1$ is assumed, which allows great simplification for mathematics. The density at $t+\delta t$ is calculated as
\begin{equation}
\begin{split}
    &\rho\left({\bf x},t+\delta t\right) =  \sum_{\beta}w_{\beta}\left[\rho\left({\bf x}-{\bf c}\delta t,t\right)+\frac{1}{2}\frac{c_{\beta i}F_{i}\left({\bf x}-{\bf c}\delta t,t\right)}{c_s^2}\delta t\right]\\
    & =\sum_{\beta}w_{\beta}\left[\rho +\frac{1}{2}\delta t^2c_{\beta j}c_{\beta k}\partial_{j}\partial_{k}\rho + \frac{1}{24}\delta t^4 c_{\beta j}c_{\beta k}c_{\beta l}c_{\beta m}\partial_{j}\partial_{k}\partial_{l}\partial_{m}\rho+\cdots\right]\\
   & - \frac{1}{2}\sum_{\beta}w_{\beta}\left[\frac{\delta t^2}{c_s^2}\partial_{j}F_{i}c_{\beta i}c_{\beta j}+\frac{\delta t^4}{6c_s^2}\partial_{j}\partial_{k}\partial_{l}F_{i}c_{\beta i}c_{\beta j}c_{\beta k}c_{\beta l}+\cdots\right],
    \end{split}
    \label{eq:secondisotropy}
    \end{equation}
    where all the odd-order terms in the Taylor series are zero due to the symmetry of the discrete velocity set. Physically, we should then require the $n$th order tensors $\sum_{\beta}w_{\beta}\underbrace{c_{\beta i}c_{\beta j}c_{\beta k}c_{\beta l}\cdots}_{n~c_{\beta}}$ to be isotropic, as otherwise the density field $\rho(t+\delta t)$ would have already contained errors due to deviations from isotropy. Having established that, it is straightforward to realize the incompleteness of Shan's remedy: without concurrently enforcing the isotropy in Eq.~(\ref{eq:secondisotropy}) to a similar order at the same time, the isotropy of a multiphase LB simulation would not be effectively improved. This is probably the reason why spurious currents were not reduced as significantly as expected with Shan's remedy,  only by a factor of 3 when the isotropy of the interaction force was increased from 4th-order to 8th-order~\cite{shan2006analysis}. 
 
 \section{A complete remedy for the lack of isotropy}
 
 \begin{table*}
\caption{Parameters in the D2Q9, D2Q13 and D2Q25 models }
\footnotesize
\begin{center}
\begin{tabular}
{c c c c c c c c c c c}
\hline
\toprule
& $c_s^2$ &$w(0)$& $w(1)$&$w(2)$&$w(3)$&$w(4)$&$w(5)$&$w(6)$&$w(7)$&$w(8)$\\
\hline
\midrule 
D2Q9& 1/3&4/9&1/9&1/36&&&&&&\\
D2Q13& 2/5&2/5&8/75&1/25&&1/300&&&&\\
D2Q25& 4/7&72/245&16/147&16/315&&1/105&8/2205&&&1/8820\\
\hline
\bottomrule
\end{tabular} 
\end{center}
\label{tab:parameters}
\end{table*}

\begin{figure}
\centerline{\includegraphics[width=160mm]{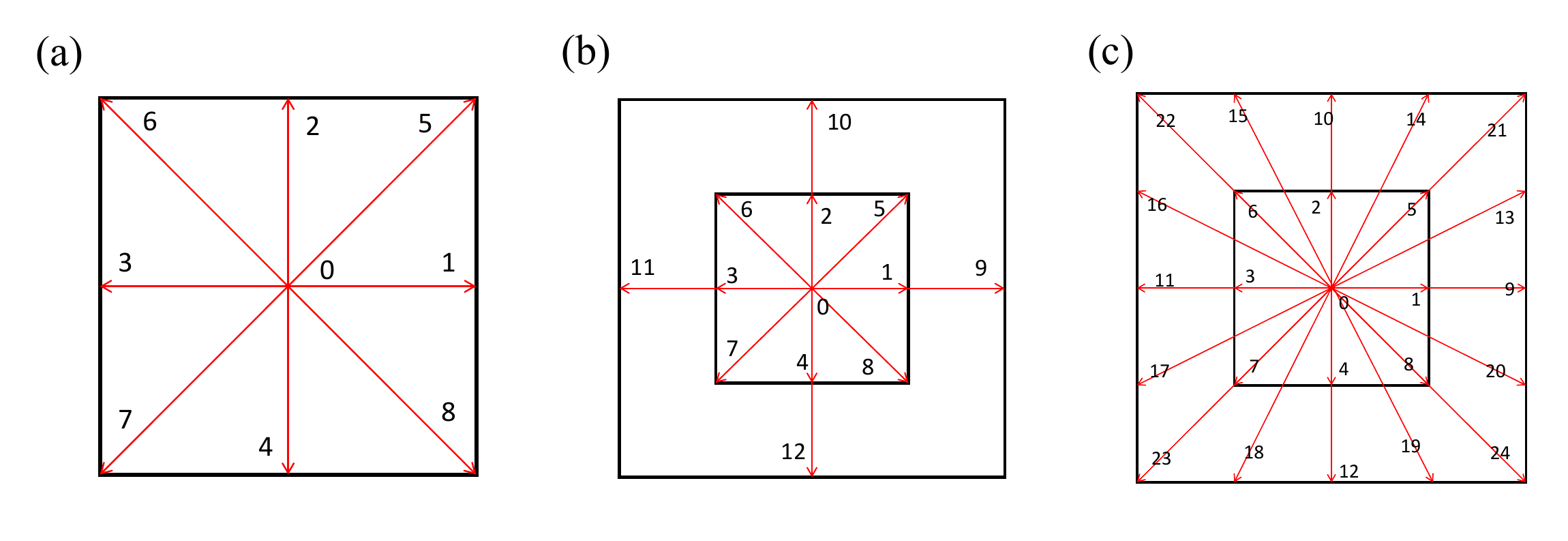}}
\vspace{0.10in}
\caption{Velocity stencils: (a) D2Q9, (b) D2Q13, (c) D2Q25. }
\label{fig:stencils}
\end{figure}

    Essentially, this second aspect of isotropy concerns the distribution of the interaction force
back to the lattice nodes, while the first aspect in Shan's analysis concerns the calculation of the interaction force. This second isotropy
    is determined by $w_{\beta}$ and ${\bf c}_{\beta}$, which could not be remedied without expanding the discrete velocity sets in the LB model. It has been proven that the nearest-neighboring LB models, such as D2Q9, D3Q15, D3Q19, and D3Q27 can only achieve a fourth-order isotropy with their discrete velocity sets~\cite{shan2006analysis}. However, with the stencils in Fig.~\ref{fig:stencils}b and Fig.~\ref{fig:stencils}c, we can easily construct a D2Q13 model and a D2Q25 model to increase the second isotropy to 6th- and 8th-order, respectively. These two models can have the same form of equilibrium distribution function and forcing function as the regular LB models, except that the weighting factor $w_{\beta}$ and the speed of sound $c_s$ have to be redefined. These parameters are given in Table~\ref{tab:parameters}. Other models allowing even higher-orders of isotropy can be formulated using the same philosophy. The velocity stencils and the corresponding weighting factors given in literature~\cite{shan2006analysis,sbragaglia2007generalized} can be used as references, the remaining job is to design the equilibrium distribution function and the forcing function {\it i.e.}, $f_{\beta}^{(eq)}$ and $\phi_{\beta}$ to satisfy the constraints that lead to the reproduction of the Navier-Stokes equation. It is worth mentioning that Mattila {\it et al.} also mentioned that LB models with larger sets of discrete velocities could be employed to reduce the level of spurious currents~\cite{mattila2013high}. However, their models were designed to incorporate higher-order time-integration schemes to replace the trapezoidal rule in standard LB models rather than to maximize the order of isotropy. Therefore, their models with the largest number of discrete velocities did not show the most significant reduction of spurious currents. The relationship between the lack of isotropy in the distribution of interaction force and the appearance of spurious currents was not explicitly stated.
    
\begin{figure}
\centerline{\includegraphics[width=160mm]{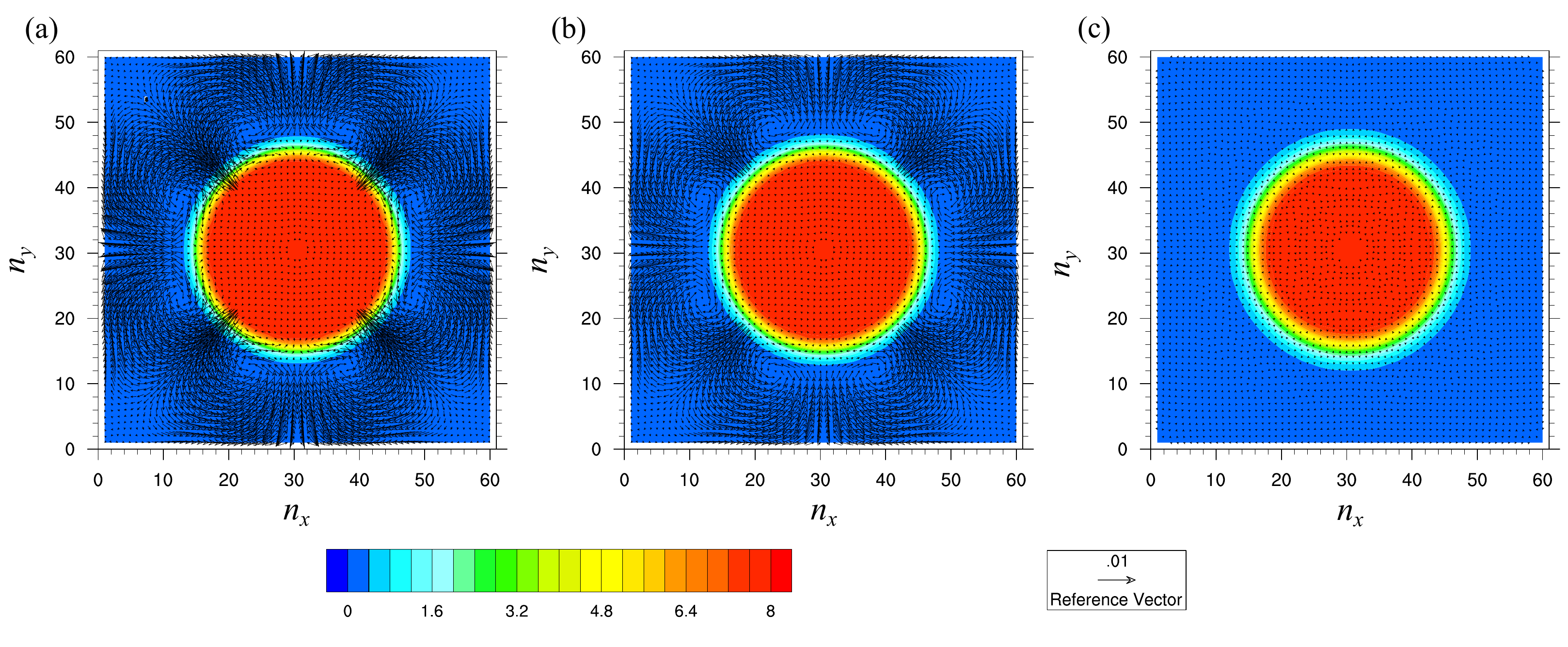}}
\caption{Density contours and velocity vectors around a stationary drop at the steady state at reduced temperature $T/T_c = 0.725$: (a) D2Q9, 4th-order isotropy, (b) D2Q9, 8th-order isotropy, (c) D2Q25. }
\label{fig:spuriouscurrents}
\end{figure}

    We employ a simple test case of a droplet suspended in a 2D periodic domain in vapor. The grid resolution of the test is $60\times60$, and a droplet with an initial radius of $r_{0} = 15$ is placed at the center of the domain $(x_c,y_c)$. The initial density distribution is defined as $\rho_{0}(x,y) = \frac{\rho_l+\rho_v}{2}-\frac{\rho_{l}-\rho_{v}}{2}\tanh{\left[\frac{2\sqrt{\left(x-x_{c}\right)^2+\left(y-y_{c}\right)^2-r_{0}}}{W}\right]}$, where $\rho_{l}$ and $\rho_{v}$
are the liquid and vapor density, respectively, at a given temperature $T$ below the critical temperature $T_{c}$. The equation of state (EOS) used in the simulations is the Peng-Robinson (P-R) EOS~\cite{peng1976new}, $p = \frac{\rho RT}{1-b\rho} - \frac{a\alpha(T)\rho^2}{1+2b\rho-b^2\rho^2}$, where $a=2/49$, $b=2/21$, $R=1$, as defined in Yuan and Schaefer~\cite{yuan2006equations}, $\alpha(T) = [1+(0.37464+1.54226\omega - 0.26992\omega^2)(1-\sqrt{T/T_c})]^2$, $\omega$ is chosen to be 0.344 for water. The pseudo-potential function $\psi$ is calculated as $\psi = \sqrt{\frac{2(p-c_s^2\rho)}{Gc_s'^2\delta t}}$, $c_s^2 = \sum_\beta w_{\beta}c_{\beta i}c_{\beta i}$, $c_s'^2 = \sum_\alpha w_{\alpha}e_{\alpha i}e_{\alpha i}$. The exact difference method (EDM)~\cite{kupershtokh2009equations} is adopted as the forcing scheme to distributed the interaction force in LB models. At a reduced temperature $T/T_c = 0.725$, the steady state density contours and velocity fields generated with the standard D2Q9 model,  the modified D2Q9 model under Shan's improvement with the eighth-order isotropy to compute the interaction force, and D2Q25 model are shown in Fig.~\ref{fig:spuriouscurrents}. Clearly, although Shan's improvement was able to reduce the magnitude of the spurious velocity to certain extent, it is not so effective compared to the D2Q25 model. More importantly, the azimuthal-dependent flow patterns still exist with Shan's best improvement, which indicates that there are remaining errors due to the lack of isotropy. The same flow patterns are no longer exist with the D2Q25 model, the remaining spurious currents are almost perpendicular to the interface. 

\begin{figure}
\centerline{\includegraphics[width=150mm]{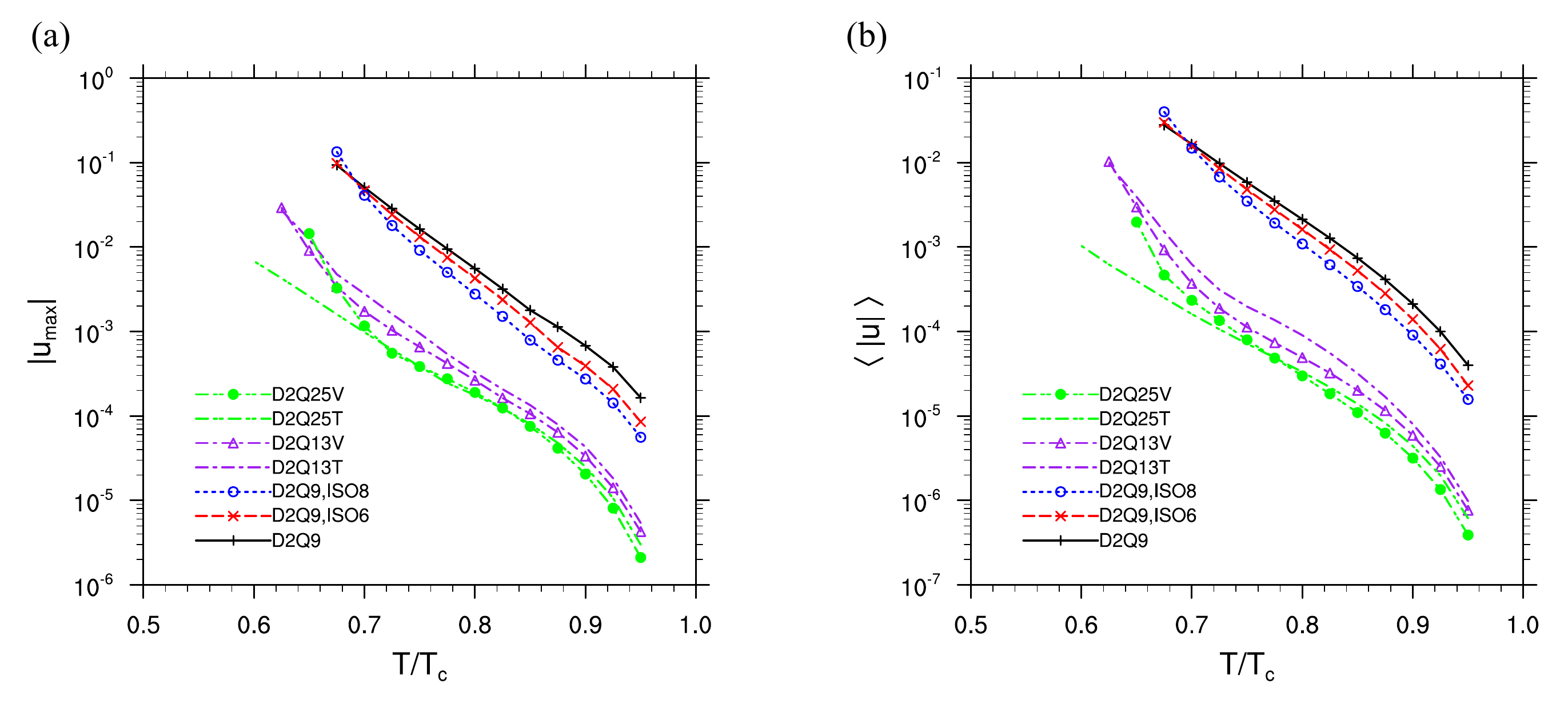}}
\caption{The magnitude of spurious velocity of a stationary drop at the steady state as a function of reduced temperature: (a) maximum spurious velocity, (b) field-averaged spurious velocity. }
\label{fig:magnitude}
\end{figure}

To quantify the reduction of spurious currents by the proposed models, we calculated the maximum spurious velocity and the field-averaged spurious velocity for a larger range of the reduced temperature. These results are shown in Fig.~\ref{fig:magnitude}. For D2Q13 and D2Q25 models, two different relaxation times are used. One is chosen identical to the relaxation time in the D2Q9 models $\tau = 1.0$, which are labeled as D2Q13T and D2Q25T. The other is designed to result in the same viscosity as in the D2Q9 models, {\it i.e.}, $\tau = 11/12$ for the D2Q13 model and $\tau = 19/24$ for the D2Q25 model. The latter two simulations are labeled as D2Q13V and D2Q25V. For the examined reduced temperature range, $0.6\le T/Tc \le 0.95$, which covers a range of liquid-to-vapor density ratio (from 4 to 800), D2Q13 and D2Q25 models always reduce the magnitude of the spurious velocity by another order of magnitude compared to Shan's improvement. It is worth noticing that Shan's improvement becomes ineffective at small reduced temperatures. As we shall see shortly, this is probably because Shan's improvement increases a thermodynamic inconsistency that offsets the benefit of increasing the isotropy in calculating the interaction force.

The thermodynamic inconsistency is another critical issue in pseudo-potential models. The lack of thermodynamic inconsistency can usually be seen from the deviations of the numerically obtained liquid and vapor densities from the corresponding values obtained by the Maxwell equal area rule, for a given EOS of a pure substance. In the literature, there are many attempts to quantify the magnitude of such derivations in the suspending droplet case shown above. However, the droplet case should not be used to evaluate such deviations from the results of the Maxwell  equal  area rule, as the curved liquid-vapor interface in this case leads to different pressures in the liquid and vapor bulk phases while the Maxwell  equal  area rule is based on a same pressure in the two phases. A more appropriate case to measure the thermodynamic inconsistency in the pseudo-potential models is a 1D flat interface case. In this case, we have measured the numerically obtained liquid and vapor densities at different reduced temperatures with P-R EOS. The new models (D2Q13, D2Q25) in general do not alter the coexisting liquid-vapor densities significantly. In fact, they slightly improve the liquid-vapor densities at smaller reduced temperatures compared to Shan's improvements (D2Q9 ISO6, D2Q9 ISO8). The larger deviations from the thermodynamic consistency with Shan's improvement may explain its failure to reduce the spurious velocity at small reduced temperatures observed in Fig.~\ref{fig:magnitude}. It is worth mentioning that there are also many available ways to improve the thermodynamic consistency in the literature, such as using a coupled form to construct the intermolecular force~\cite{kupershtokh2009equations}, modifying EOS~\cite{colosqui2012mesoscopic}, and introducing correction terms in the forcing schemes of LBM~\cite{li2012forcing}. These available improvements can be used to improve the thermodynamic consistency in our models. Therefore, we do not address the issue of thermodynamic inconsistency in the present study.

\begin{figure}
\centerline{\includegraphics[width=90mm]{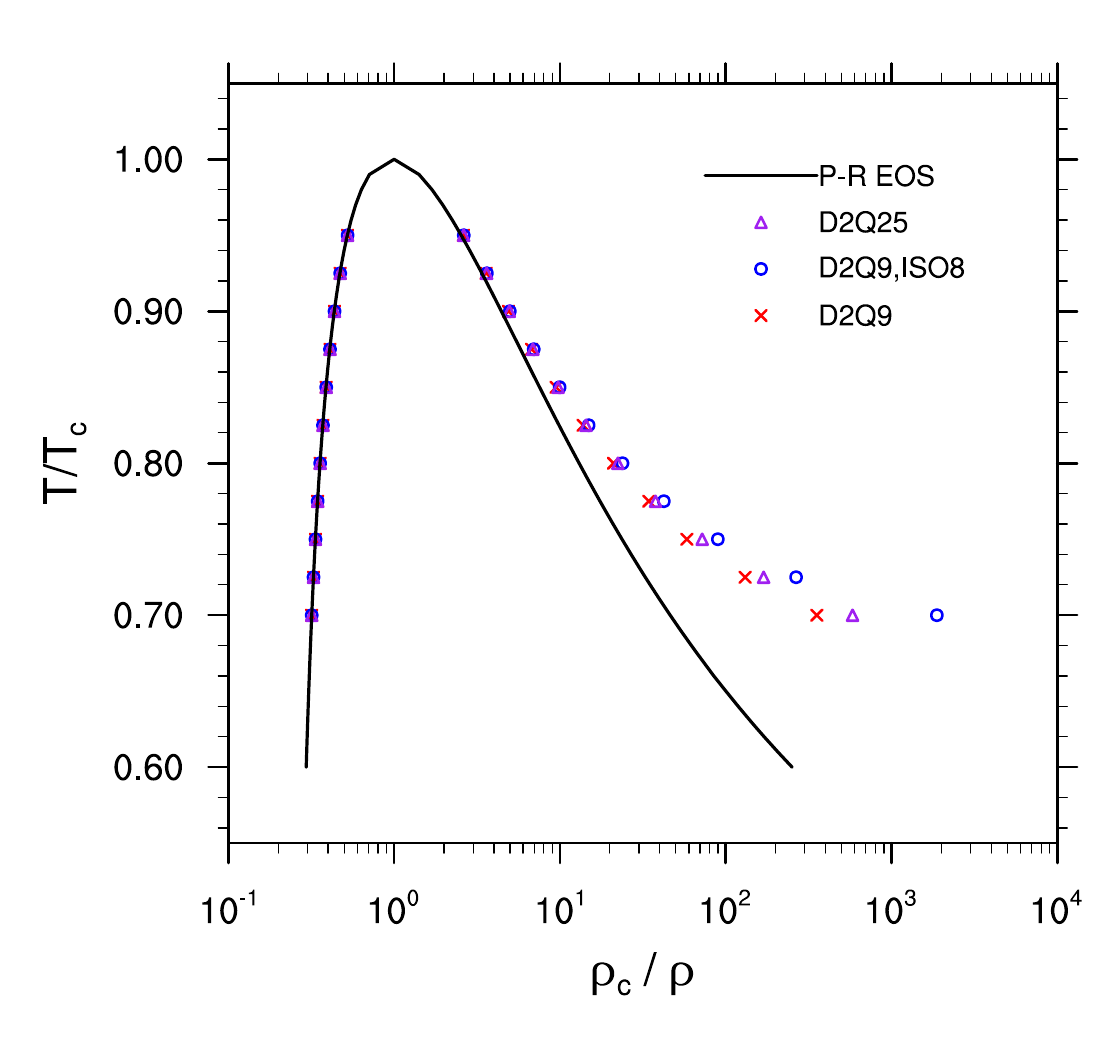}}
\caption{The coexisting liquid-vapor densities as functions of reduced temperature in a flat interface test. }
\label{fig:Coexistence}
\end{figure}

\begin{figure}
\centerline{\includegraphics[width=90mm]{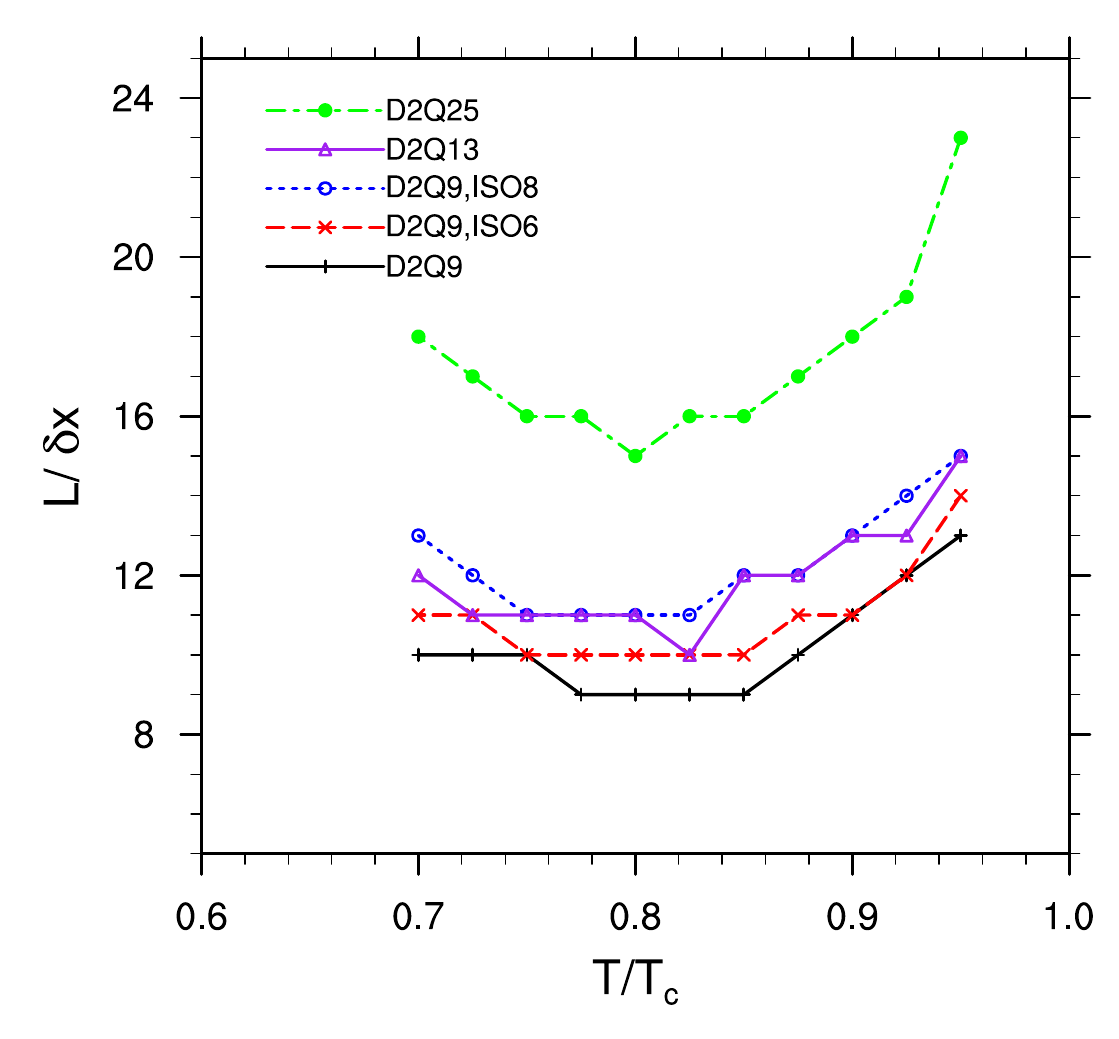}}
\caption{The Interface thicknesses as functions of reduced temperature in a flat interface test. }
\label{fig:Thickness}
\end{figure}

The major side effect of using D2Q13 and D2Q25 modes is the increased interface thickness. In the 1D flat interface case, the interface thickness, defined as the region with $1.05\rho_{v}\le \rho \le 0.95\rho_{l}$, where $\rho_{v}$ and $\rho_{l}$ are the numerically obtained vapor and liquid densities in the two-phase bulk regions with each model, are measured and shown in Fig.~\ref{fig:Thickness}. The increased interface thickness comes from two aspects. First, by using large stencils to compute the interaction force in the pseudo-potential models, the local force depends on the potential $\psi$ from more neighboring grid points, which makes the interface broader. This aspect also impacts the interface thickness when using Shan's improvement, but it has not been emphasized in the literature. Second, the use of large lattice models allows a local grid point to directly communicate with not just the nearest neighboring nodes, but also the next layer of neighboring nodes, which adversely affects the locality of LBM. This aspect also makes the interface thicker. If the thickening interface has to be avoided in a certain application, the D2Q13 model could be a better choice compared to the D2Q25 model.

Finally, we would like to briefly comment on the no-slip boundary treatment for the proposed models. In general, the no-slip boundary can still be treated following the bounce-back schemes. However, the uses of certain bounce-back schemes, such as the half-way bounce-back may not be straightforward, as it is difficult to place a solid wall precisely half-way for all links. On the other hand, schemes such as the standard bounce-back~\cite{noble1995direct} and modified bounce-back~\cite{inamuro1995non} are not affected. Additional attention to be paid is that the unknown boundary distribution functions have to be constructed on the first two layers of interior grid points. As a demonstration, a test case of a droplet contacting with a flat wall has been added. The P-R EOS is still used and the temperature is set to $T = 0.9T_c$. The standard bounce-back scheme is adopted to enforce the no-slip condition. The varying wettability in the two new models can still be achieved by tuning the virtual density of the wall $\rho_w$ as in Ref.~\cite{sbragaglia2006surface}.

\begin{figure}
\centerline{\includegraphics[width=160mm]{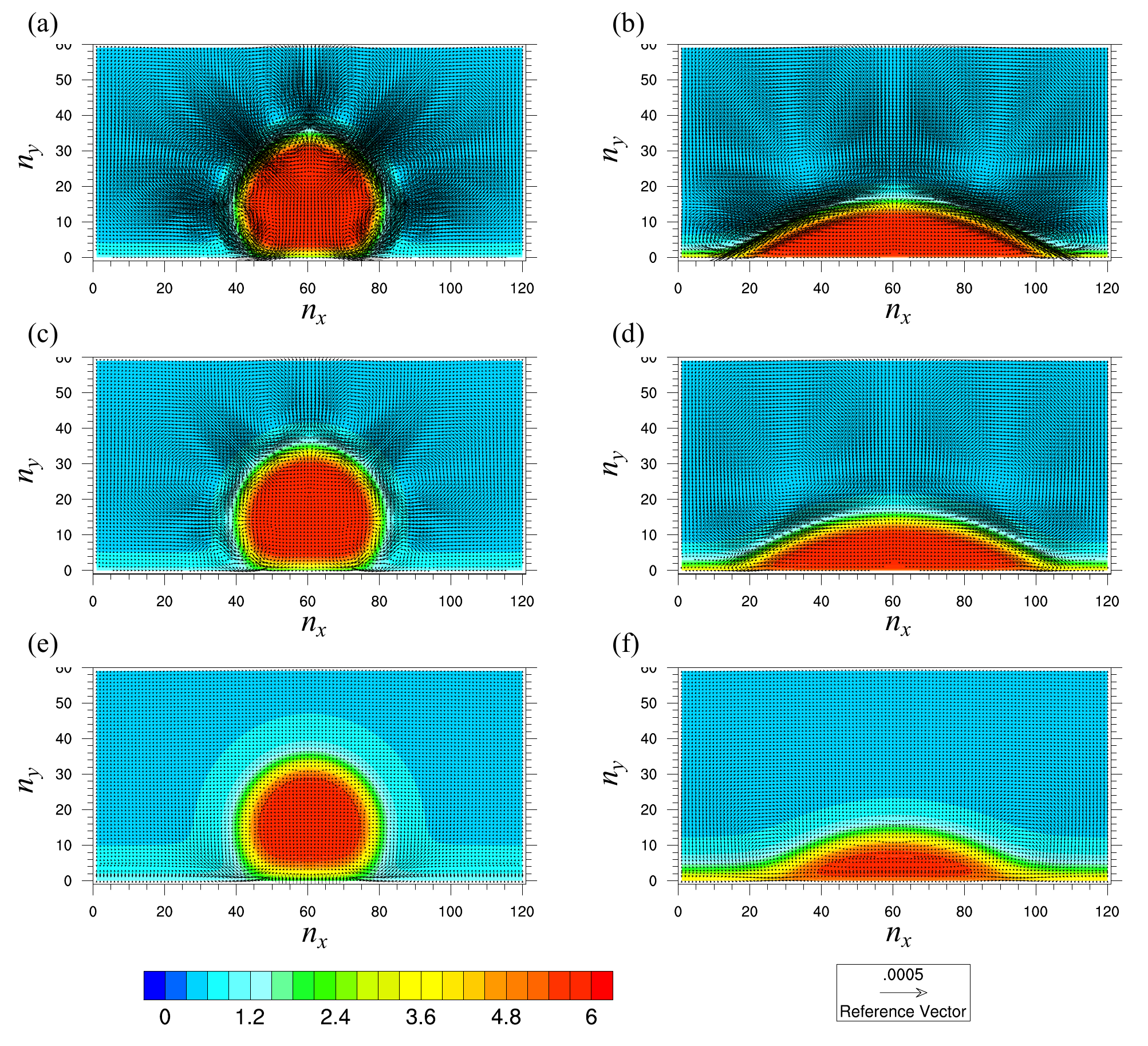}}
\caption{Density contours and velocity vectors in a droplet contacting with a flat wall: (a) D2Q9, 4th-order isotropy, $\rho_{w} = 1.5$ (b) D2Q9, 4th-order isotropy, $\rho_{w} = 4.5$, (c) D2Q9, 8th-order isotropy, $\rho_{w} = 1.5$, (d) D2Q9, 8th-order isotropy, $\rho_{w} = 4.5$, (e) D2Q25, $\rho_{w} = 1.5$, (f) D2Q25, $\rho_{w} = 4.5$.}
\label{fig:Contact}
\end{figure}

\section{Conclusion and discussion}
In this work, we point out the incompleteness of the previous understanding of the spurious currents in the multiphase flow simulation with the pseudo-potential LB models. There are two types of isotropy requirements: the first concerns the macroscopic force calculation, and the second is the mesoscopic redistribution. The two should be considered together in order to more effectively reduce spurious currents. We proposed two LB models with more discrete velocities that can reduce spurious currents by one order of magnitude.

The remaining spurious velocities are associated with the discrete nature in LB models. Following the same analysis in Eq.~(\ref{eq:secondisotropy}), the momentum at $({\bf x}, t+\delta t)$ can be computed as
\begin{equation}
\begin{split}
   & \rho u_{i}\left({\bf x}, t+\delta t\right) \\
   &=  \sum_{\beta}w_{\beta}\left[\rho\left({\bf x}-{\bf c}\delta t,t\right)+\frac{1}{2}\frac{c_{\beta j}F_{j}\left({\bf x}-{\bf c}\delta t,t\right)}{c_s^2}\delta t\right]c_{\beta i} + \frac{1}{2}F_{i}\left({\bf x}, t\right)\delta t\\
   & = \sum_{\beta}\left(-w_{\beta}c_{\beta i}c_{\beta j}\delta t\partial_{j}\rho - \frac{1}{6}c_{\beta i}c_{\beta j}c_{\beta k}c_{\beta l}\delta t^3\partial_{j}\partial_{k}\partial_{l}\rho+ ...\right)\\
   &+\frac{1}{2}\sum_{\beta}\left(\frac{\delta tF_{j}}{c_s^2}w_{\beta}c_{\beta i}c_{\beta j}+\frac{\delta t^3\partial_{k}\partial_{l}F_{j}}{2c_s^2}c_{\beta i}c_{\beta j}c_{\beta k}c_{\beta l} + \cdots\right) + \frac{1}{2}F_{i}\delta t\\
   & = \delta t\left(F_{i} - c_s^2\partial_{i}\rho \right) + \delta_{t}^3 \frac{3c_{s}^2}{4}\partial_{k}\partial_{k}\left(F_{i}-\frac{2c_{s}^2}{3}\partial_{i}\rho\right) + ... 
    \end{split}
    \label{eq:theory}
\end{equation}
Regardless of whether the truncated high-order tensors are isotropic or not, $\rho u_{i}({\bf x},t+\delta t)$ appears always non-zero since the first two terms cannot be zero at the same time. Even if we only keep the leading-order term in Eq.~(\ref{eq:theory}), the coefficient $\delta t F_{i}/c_s^2-\delta t\partial_{j}\rho\delta_{ij}$ may not be precisely zero. Ideally, on the N-S equation level this coefficient should vanish when a hydrostatic balance is established, but once discretized, the precise balance is usually violated. The spurious currents due to this imbalance are aligned along with the direction of pressure gradient, which corresponds to what we observed in Fig.~\ref{fig:spuriouscurrents}c. On a 1D flat interface, this spurious current further reduces to what Guo {\it et al}. reported as inevitable artificial velocities in LB simulations~\cite{guo2011force}. The pseudo-potential LB models calculate the interaction force in a discretized form and again distribute this force to the discrete distribution functions, both inducing errors that could result in spurious currents. A consistent consideration of the two processes together may help further suppress or even remove the spurious current, which will be pursued in the future.

{\bf Acknowledgements}: Authors would like to thank Prof. Zhaoli Guo at Huazhong University of Science and Technology, China, and Prof. Haibo Huang at University of Science and Technology of China for the inspiring discussions. Funding support from Energi Simulation and the William A. Fustos Family Professorship in Energy and Mineral Engineering at Penn State University are gratefully acknowledged.



\bibliographystyle{elsarticle-num}

\bibliography{elsarticle-template}

\begin{thebibliography}{10}
\expandafter\ifx\csname url\endcsname\relax
  \def\url#1{\texttt{#1}}\fi
\expandafter\ifx\csname urlprefix\endcsname\relax\def\urlprefix{URL }\fi
\expandafter\ifx\csname href\endcsname\relax
  \def\href#1#2{#2} \def\path#1{#1}\fi

\bibitem{shan1993lattice}
X.~Shan, H.~Chen, Lattice boltzmann model for simulating flows with multiple
  phases and components, Physical Review E 47~(3) (1993) 1815.

\bibitem{shan1994simulation}
X.~Shan, H.~Chen, Simulation of nonideal gases and liquid-gas phase transitions
  by the lattice boltzmann equation, Physical Review E 49~(4) (1994) 2941.

\bibitem{chen2014critical}
L.~Chen, Q.~Kang, Y.~Mu, Y.-L. He, W.-Q. Tao, A critical review of the
  pseudopotential multiphase lattice boltzmann model: Methods and applications,
  International Journal of Heat and Mass Transfer 76 (2014) 210--236.

\bibitem{li2016lattice}
Q.~Li, K.~H. Luo, Q.~Kang, Y.~He, Q.~Chen, Q.~Liu, Lattice boltzmann methods
  for multiphase flow and phase-change heat transfer, Progress in Energy and
  Combustion Science 52 (2016) 62--105.

\bibitem{wagner2003origin}
A.~J. Wagner, The origin of spurious velocities in lattice boltzmann,
  International Journal of Modern Physics B 17~(01n02) (2003) 193--196.

\bibitem{shan2006analysis}
X.~Shan, Analysis and reduction of the spurious current in a class of
  multiphase lattice boltzmann models, Physical Review E 73~(4) (2006) 047701.

\bibitem{lee2006eliminating}
T.~Lee, P.~F. Fischer, Eliminating parasitic currents in the lattice boltzmann
  equation method for nonideal gases, Physical Review E 74~(4) (2006) 046709.

\bibitem{yuan2006equations}
P.~Yuan, L.~Schaefer, Equations of state in a lattice boltzmann model, Physics
  of Fluids 18~(4) (2006) 042101.

\bibitem{yu2010multirelaxation}
Z.~Yu, L.-S. Fan, et~al., Multirelaxation-time interaction-potential-based
  lattice boltzmann model for two-phase flow, Physical Review E 82~(4) (2010)
  046708.

\bibitem{guo2011force}
Z.~Guo, C.~Zheng, B.~Shi, Force imbalance in lattice boltzmann equation for
  two-phase flows, Physical Review E 83~(3) (2011) 036707.

\bibitem{guo2002discrete}
Z.~Guo, C.~Zheng, B.~Shi, Discrete lattice effects on the forcing term in the
  lattice {B}oltzmann method, Physical Review E 65~(4) (2002) 046308.

\bibitem{sbragaglia2007generalized}
M.~Sbragaglia, R.~Benzi, L.~Biferale, S.~Succi, K.~Sugiyama, F.~Toschi,
  Generalized lattice boltzmann method with multirange pseudopotential,
  Physical Review E 75~(2) (2007) 026702.

\bibitem{mattila2013high}
K.~K. Mattila, D.~N. Siebert, L.~A. Hegele~Jr, P.~C. Philippi, High-order
  lattice-boltzmann equations and stencils for multiphase models, International
  Journal of Modern Physics C 24~(12) (2013) 1340006.

\bibitem{peng1976new}
D.-Y. Peng, D.~B. Robinson, A new two-constant equation of state, Industrial \&
  Engineering Chemistry Fundamentals 15~(1) (1976) 59--64.

\bibitem{kupershtokh2009equations}
A.~Kupershtokh, D.~Medvedev, D.~Karpov, On equations of state in a lattice
  boltzmann method, Computers \& Mathematics with Applications 58~(5) (2009)
  965--974.

\bibitem{colosqui2012mesoscopic}
C.~E. Colosqui, G.~Falcucci, S.~Ubertini, S.~Succi, Mesoscopic simulation of
  non-ideal fluids with self-tuning of the equation of state, Soft matter
  8~(14) (2012) 3798--3809.

\bibitem{li2012forcing}
Q.~Li, K.~H. Luo, X.~Li, et~al., Forcing scheme in pseudopotential lattice
  boltzmann model for multiphase flows, Physical Review E 86~(1) (2012) 016709.

\bibitem{noble1995direct}
D.~R. Noble, J.~G. Georgiadis, R.~O. Buckius, Direct assessment of lattice
  boltzmann hydrodynamics and boundary conditions for recirculating flows,
  Journal of statistical physics 81~(1-2) (1995) 17--33.

\bibitem{inamuro1995non}
T.~Inamuro, M.~Yoshino, F.~Ogino, A non-slip boundary condition for lattice
  boltzmann simulations, Physics of Fluids 7~(12) (1995) 2928--2930.

\bibitem{sbragaglia2006surface}
M.~Sbragaglia, R.~Benzi, L.~Biferale, S.~Succi, F.~Toschi, Surface
  roughness-hydrophobicity coupling in microchannel and nanochannel flows,
  Physical review letters 97~(20) (2006) 204503.

\end{thebibliography}

\end{document}